\documentclass{emulateapj}
\begin{document}

\title{Initial Size Distribution of the Galactic Globular Cluster System}
\author{
Jihye Shin\altaffilmark{1},
Sungsoo S. Kim\altaffilmark{1,2},
Suk-Jin Yoon\altaffilmark{3}, and
Juhan Kim\altaffilmark{4}}
\email{jhshin@ap4.khu.ac.kr}
\altaffiltext{1}{Department of Astronomy \& Space Science,
Kyung Hee University, Yongin, Kyungki
446-701, Republic of Korea}
\altaffiltext{2}{School of Space Research, Kyung Hee University,
Yongin, Kyungki 446-701, Republic of Korea}
\altaffiltext{3}{Department of Astronomy and Center for Galaxy Evolution
Research, Yonsei University, Seoul 120-749, Republic of Korea}
\altaffiltext{4}{Center for Advanced Computation, Korea Institute for Advanced
Study, 87 Hoegiro Dondaemun-gu, Seoul 130-722, Republic of Korea}

\begin{abstract}
Despite the importance of their size evolution in understanding the
dynamical evolution of globular clusters (GCs) of the Milky Way,
studies are rare that focus specifically on this issue. Based on the
advanced, realistic Fokker--Planck (FP) approach, we predict theoretically
the initial size distribution (SD) of the Galactic GCs along with their
initial mass function and radial distribution. Over one thousand FP
calculations in a wide parameter space have pinpointed the best-fit
initial conditions for the SD, mass function, and radial distribution.
Our best-fit model shows that the initial SD of the Galactic GCs is
of larger dispersion than today's SD, and that typical projected half-light
radius of the initial GCs is $\sim$4.6~pc, which is 1.8 times larger
than that of the present-day GCs ($\sim$2.5~pc). Their large size
signifies greater susceptibility to the Galactic tides: the total mass
of destroyed GCs reaches 3--5$\times$$10^8~M_{\odot}$, several times
larger than the previous estimates. Our result challenges a recent
view that the Milky Way GCs were born compact on the sub-pc scale, and
rather implies that (1) the initial GCs are generally larger than
the typical size of the present-day GCs, (2) the initially large GCs
mostly shrink and/or disrupt as a result of the galactic tides, and (3)
the initially small GCs expand by two-body relaxation, and later shrink
by the galactic tides.
\end{abstract}
\keywords{Galaxy: evolution - Galaxy: formation - Galaxy:
kinematics and dynamics - globular clusters: general - methods: numerical}

\section{Introduction}
Whereas the present-day mass functions (MFs) of globular cluster (GC)
systems, which are nearly universal among galaxies \citep{bro06,jor07},
are approximately log-normal with a peak mass
$M_{p}\approx2$$\times$$10^{5}~M_{\odot}$, the MFs of the young massive
star cluster (YMC) systems follow a simple power-law distribution
\citep[among others]{whi95,zha99,deg03}. Motivated by such a difference
between GCs and YMCs, numerous studies have examined the dynamical evolution
of the GC MFs to determine whether the initial MFs of GC systems resemble
those of YMC systems \citep[among others]{gne97,bau98,ves98,fal01,par07,shi08}.
In particular, \citet[Paper I hereafter]{shi08} surveyed a wide range of
parameter space for the initial conditions of the Milky Way GCs, and
considered virtually all internal/external processes: two-body relaxation,
stellar evolution, binary heating, galactic tidal field, eccentric orbits
and disc/bulge shocks. They found that the initial GC MF that best fits
the observed GC MF of the Milky Way is a log-normal function with a peak at
4$\times$$10^5~M_{\odot}$ and a dispersion of 0.33, which is quite different
from the typical MFs of YMCs.

Using the outcome of $N$-body calculations, \citet{gie08} found that the aspect
of mass loss in GCs varies with the tidal filling ratio
$\Re \equiv r_{h}/r_{J}$, where $r_h$ is the half-mass radius and $r_{J}$
is the Jacobi radius. More specifically, the mass loss of GCs in the
"isolated regime" ($\Re < 0.05$) is driven mostly by the two-body
relaxation, which induces the formation of binaries in the core and
causes GCs to expand. On the other hand, the mass loss of GCs in the
"tidal regime" ($\Re > 0.05$) is influenced by the galactic tides
as well, which enables stars in the outer envelope to easily escape
(evaporation). Thus, the cluster size ($r_h$) is as important as the
cluster mass ($M$) and the galactocentric radius ($R_G$) in determining
the dynamical evolution of GCs.

Can YMCs tell us something about the typical initial size of the Milky Way
GC system? Observations show that the projected half-light radius
$R_h$ of YMCs (ages up to 100 Myr) in the local group ranges between
$\sim$2 and $\sim$30~pc with a mean value of $\sim$8~pc \citep{pz10},
which is a few times larger than that of the present-day Milky Way GCs,
$R_h\sim2.5$~pc.
However, GCs could have formed in different environments and/or by different
mechanisms from the YMCs.

Perhaps the best way to estimate the typical size of the GCs is to trace
them back to their initial state by calculating their dynamical evolution.
In this paper, we study the dynamical evolution of the Galactic GCs and
identify the most probable initial conditions not only for the MF and
radial distribution (RD), but also the size distribution (SD). Using the
same numerical method and procedure as in Paper I, we perform Fokker-Planck
(FP) calculations for 1152 different initial conditions (mass, half-mass
radius, galactocentric radius and orbit eccentricity), and then search a
wide-parameter space for the most probable initial distribution models
that evolve into the present-day Galactic GC distributions.

\begin{figure}
\includegraphics[scale=1.0,clip=true]{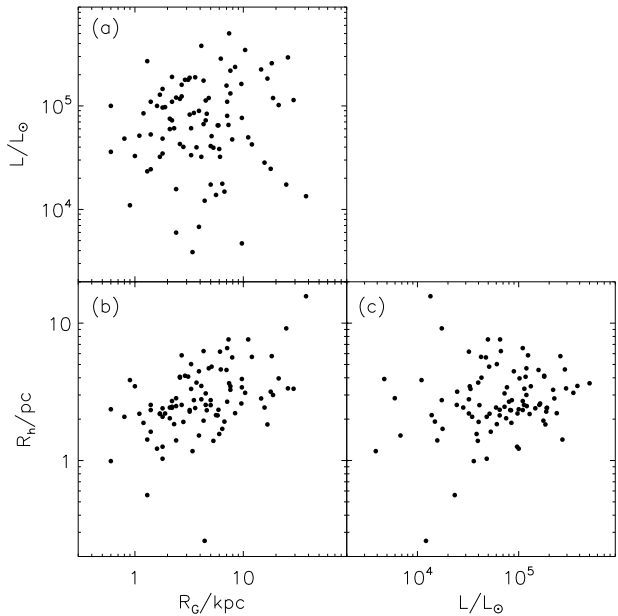}
\caption{Distribution of the Galactic "native" (see the text for definition)
globular clusters in the $L$--$R_G$ space (a), $R_h$--$R_G$ space (b), and
$R_h$--$L$ (c).  Data are from the compilation by Harris (1996).\label{obsgcs}}
\end{figure}

The paper is organized as follows. Section 2 describes the properties of
the observed GCs, again which we compare our model results. Section 3
presents models and initial conditions for FP calculations, and Section 4
analyzes the aspects of the size evolution of GCs. We synthesize our FP
results in Section 5 to construct the GC system, and examine common
features of the best-fit MF, RD, and SD models in Section 6. We discuss
characteristics of the final best-fit SD models of the Galactic GCs in
Section 7. Finally, conclusions are presented in Section 8.

\section{Present-day GC Properties}

When comparing FP calculations to the present-day Galactic GCs, we consider
the "native" GCs only, i.e., "old" halo and bulge/disc clusters, which are
believed to be created when a protogalaxy collapses while "young" halo
clusters are thought to be formed in external satellite galaxies
\citep{zin93,par00,mac05}. Our native GC candidates do not include six objects
that belong to the Sagittarius dwarf, seven objects whose origins remain
unknown, two objects that have no size information, and fifteen objects that
are thought to be the remnants of dwarf galaxies \citep{lee07}. The total
number of our present-day Galactic native GCs is 93, and their observed
properties, such as luminosity \emph{L}, $R_{h}$, and $R_G$, were obtained
from the database compiled by \citet{har96}.

Figure~\ref{obsgcs} shows scatter plots between observed \emph{L}, $R_{h}$,
and $R_G$ values for the 93 Galactic native GCs. The \emph{L}, $R_{h}$, and
$R_G$ values range between 3.9$\times$$10^3$--5.0$\times$$10^5~L_{\odot}$,
0.3--16~pc, and 0.6--38~kpc, where the mean values are located at
7.2$\times$$10^4~L_{\odot}$, 2.5~pc, and 4.1~kpc, respectively. The
correlation between $R_h$ and $R_G$ is tighter than the other two
correlations (see Figure \ref{obsgcs}$b$). This tight $R_h$--$R_G$ correlation
could be just a result of the initially tight correlation between $R_h$ and
$R_G$, or it could be due to the preferred disruption of large GCs near the
Galactic center \citep{ves97,bau03}.  Another possible cause is the expansion
of initially small GCs up to $r_J$, which is roughly proportional to
$R_G^{2/3}$ for a given GC mass.  One of the goals of this paper is to
determine which of these possibilities is more feasible.

Previous studies on the evolution of the GC system assumed a certain
constant mass-to-light (\emph{M/L}) ratio, and converted the observed $L$ to
$M$ when comparing their numerical values with observations. But the
conversion of GC luminosity function (LF) to GC MF using a constant
$\emph{M/L}$ ratio may lead to MFs in error because low-mass stars,
which have higher $M/L$ ratios than the high-mass stars,
preferentially evaporate from the cluster and this causes the $M/L$ ratio of
the cluster to evolve with time \citep{kru09}.
For the same reason, there is not a linear relationship between $R_h$
and $r_h$ among different GCs. Thus, we transform $M$ to $L$, instead of $L$ to
$M$, using the stellar mass--luminosity relation of the Padova model
\citep{mar08} with a metallicity of $[\textrm{Fe/H}]=-1.16$, which is the
mean value for the Galactic native GCs. Our FP calculations, which will be
described later, show that the present-day GCs can have $M/L$ ratios ranging
between 1.2 and 2.5 and $r_h/R_h$ ratios ranging between 1.0 and 2.5.

We use dynamical properties such as \emph{M} and $r_{h}$ when
constructing the initial distributions of the Galactic GC system and when
calculating the dynamical evolution, while observed quantities, \emph{L}
and $R_{h}$, are used when comparing our FP results with the observations.

\section{Models and Initial Conditions}

\begin{figure}
\includegraphics[scale=1.0,clip=true]{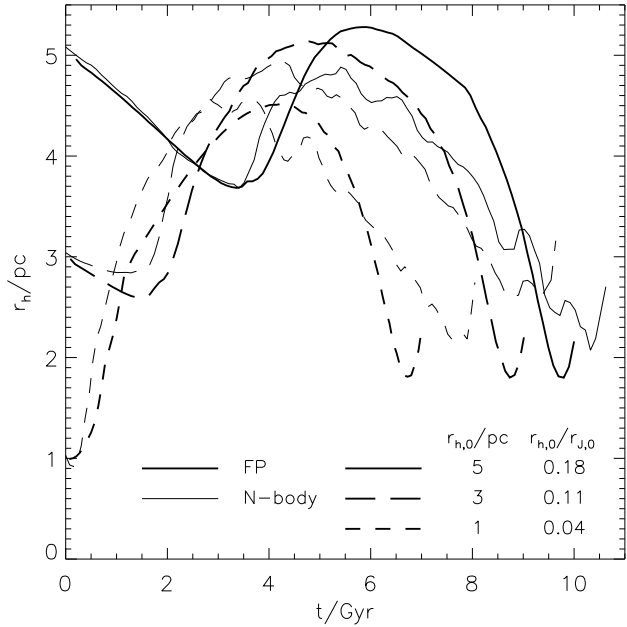}
\caption{Comparison of $r_h$ evolution between $N$-body simulations
and our FP calculations\label{nbody} for GCs with initial conditions of
$M=10^4~M_{\odot}$, $R_G=8.5$~kpc, and $r_h=$~1, 3, and 5 pc on circular
orbits. $N$-body simulations were performed using Nbody4 code \citep{aar03},
and mass loss by stellar evolution was not considered in these test
calculations (both $N$-body and FP). $r_h$ values of the two models agree
well within $\sim 20\%$ during the entire cluster lifetimes.}
\end{figure}

We adopt the anisotropic FP model used in Paper I, which was originally
developed by \citet[and references therein]{tak00}. The model integrates the
orbit-averaged FP equation of two (energy-angular momentum) dimensions and
considers multiple stellar mass components, three-body and tidal-capture
binary heating, stellar evolution, tidal fields, disk/bulge shocks, dynamical
friction, and realistic (eccentric) cluster orbit (see \citet{kim99} for the
tidal binary heating and Paper I for the detailed implementation of dynamical
friction and realistic orbits). The model implements the Alternating Direction
Implicit (ADI) method developed by \citet{shi07} for integrating the
two-dimensional FP equation with better numerical stability.

\begin{figure*}
\includegraphics[scale=0.9]{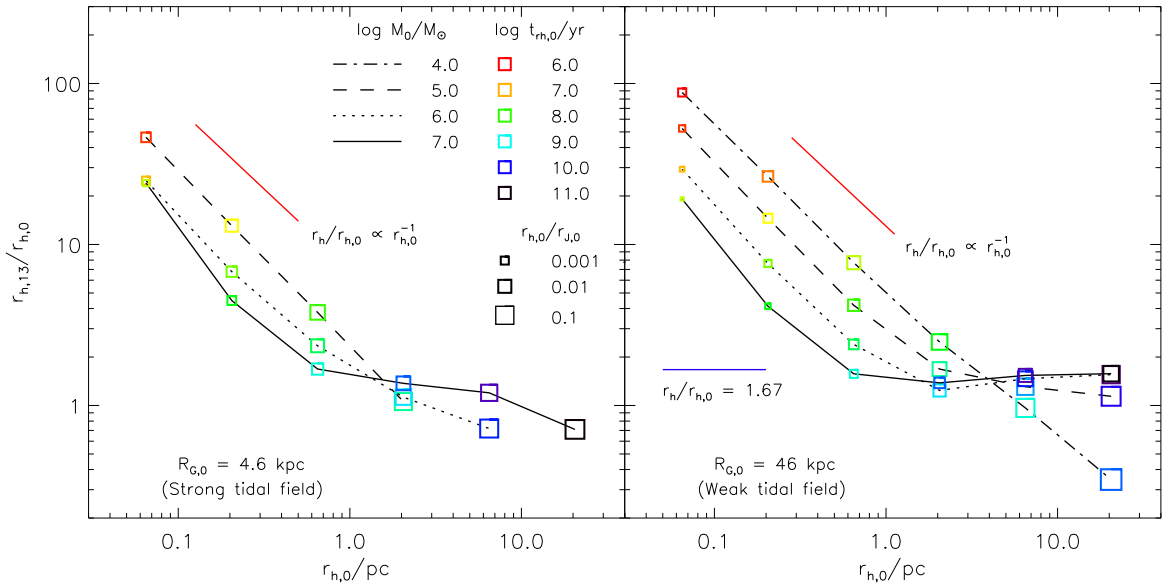}
\caption{Ratios of $r_h$ values at 13~Gyr and at the beginning from some of our
1,152 Fokker--Planck calculations as a function of $r_h$ at the beginning
for two different $R_{G,0}$ values (4.6~kpc for the left panel and 46~kpc
for the right panel) and four different initial mass ($\log M_0/M_{\odot} = 4$,
5, 6, and 7).  The approximate initial half-mass relaxation times and
the initial tidal filling ratios are marked with different colors and
symbol sizes, respectively.
The blue line indicates the location of $r_h/r_{h,0}=1.67$, which is
the expected expansion ratio mainly by the stellar evolution, and
the red lines represents the relation $r_h/r_{h,0} \propto r_{h,0}^{-1}$,
which is the expected result when the evolution is dominated by the two-body
relaxation.\label{FPresult}}
\end{figure*}
\begin{figure*}
\includegraphics[scale=1.0]{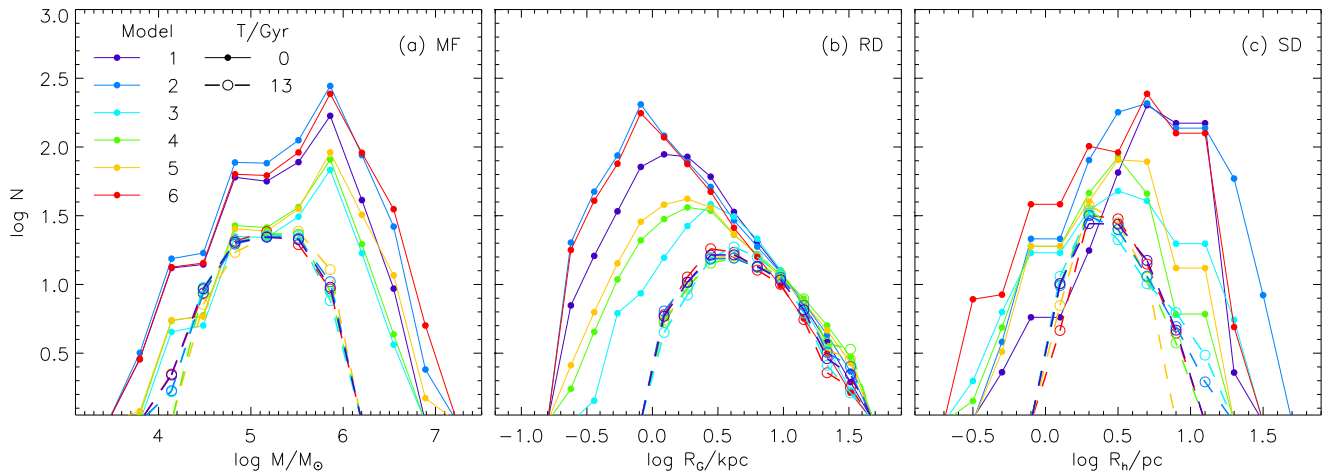}
\caption{Comparison of mass functions (a), radial distributions (b), and
size distributions (c) at 13~Gyr (solid lines) and at the beginning (dashed
lines) from the best-fit initial parameter set for each SD model.\label{mfrdsd}}
\end{figure*}

Parameters for our FP survey are the following four initial cluster
conditions: \emph{M}, \emph{$r_{h}$}, apocenter distance of the cluster orbit
\emph{$R_{a}$}, and cluster orbit eccentricity \emph{e}. We choose eight
\emph{M} values from $10^{3.5}$ to $10^{7}~M_{\odot}$, six \emph{$r_{h}$}
values from $10^{-1}$ to $10^{1.5}$~pc, and six \emph{$R_{a}$} values from
$10^{0}$ to $10^{1.67}$~kpc, all equally spaced on the logarithmic scale.
For the eccentricity, we choose \emph{e} = 0, 0.25, 0.5, and 0.75.
We perform FP calculations for all possible combinations of these four
parameters, thus the total number of cluster models considered in the present
study amounts to 1152.

For the initial stellar mass function (IMF) within each cluster, we adopt the
model developed by Kroupa (2001) with a mass range of 0.08--15~M$_{\odot}$,
which is realized by 15 discrete mass components in our FP model. Each mass
component follows the stellar evolution recipe described by \citet{sch92}.
The stellar density and velocity dispersion distributions within each cluster
follow the King model \citep{kin66} with a concentration parameter $W_{0}=7$
and with neither initial velocity anisotropy nor initial mass segregation. We
use only one value for $W_0$, thus the tidal cut-off radius $r_t$ of the King
profile is proportional to $r_h$, while $r_J$ varies depending on $M$ and
$R_G$. Therefore, the Roche lobe filling ratio ($r_t/r_J$) and $\Re$ of
our FP models are functions of $r_h$, $M$, and $R_G$.

The aspects of mass and size evolution from our FP model are in a good
agreement with those from $N$-body methods. A comparison of mass evolution
between our FP calculations and the $N$-body simulations performed
by \citet{bau03} for clusters on eccentric orbits with initial masses larger
than $10^{4}~M_{\odot}$ shows good agreement of cluster lifetimes within
$\sim$25$\%$.  For a comparison of size evolution, we run a set of $N$-body
simulations using Nbody4 code \citep{aar03} with $M=10^4~M_{\odot}$,
$R_G=8.5$~kpc, and $r_h=$~1, 3, and 5~pc (these correspond to
$\Re=$~0.04, 0.11, and 0.18), and find that the $r_h$ evolutions of the
two models agree well within $\sim 20\%$ during the entire cluster
lifetimes (see Figure~\ref{nbody}).

Due to the expulsion of the remnant gas from star formation in the
pre-gas-expulsion cluster, some of the low-mass pre-gas expulsion clusters can
quickly disrupt, and even the surviving low-mass pre-gas-expulsion clusters
will lose a significant fraction of their mass within the first several Myr and
rapidly expand \citep{bau07,par07}. Since our FP model does not consider the
effect of gas expulsion, our initial GC models are to be regarded as models at
several Myr after cluster formation.

\section{Size Evolution of Individual Globular Clusters}

The three main drivers of GC size evolution are the two-body relaxation,
the mass loss by stellar evolution, and the galactic tides.  In this section,
we discuss the size evolution of individual GCs with a subset of our FP
calculations.  Figure~\ref{FPresult} shows the ratios between $r_{h}$ values
at the present time (13 Gyr) and at the beginning from our FP calculations as
a function of $r_{h,0}$ and $M_0$ for two different $R_{G,0}$ values
(subscripts 0 denote the initial value, hereafter).

Two-body relaxation causes GC core to collapse and the subsequent
formation of dynamical binaries in the core makes the whole cluster expand.
For GCs that have undergone core collapse in the early phase of evolution,
the size of the post-core-collapse expansion follows a scaling
relation $r_h \propto M_0^{-1/3} t^{2/3}$ \citep{goo84,kim98,bau02}, and
thus for a given initial mass and epoch, $r_h/r_{h,0}$ is simply proportional
to $r_{h,0}^{-1}$. Figure~\ref{FPresult} indeed shows that the size of the GCs
with the same $M_0$ tend to converge to a single value
($r_{h}/r_{h,0}\propto r_{h,0}^{-1}$), if the GCs have small $t_{rh,0}$
($\log t_{rh,0}/$yr~$ \lesssim 9$).

Mass loss by stellar evolution causes GCs to adiabatically expand to maintain
virialization, and the GC sizes evolve following $r_h/r_{h,0}\propto M_0/M$
when the stellar evolution is the main driver of the GC size evolution
\citep{hil80}. The combination of Kroupa IMF and the stellar evolution recipe
described by \citet{sch92} yields a mass loss of $\sim40\%$ within 13~Gyr.
Thus, GCs would expand by a factor of $\sim1.67$ as a result of the stellar
evolution, if two-body relaxation or the galactic tides are relatively less
important in driving the size evolution. Indeed, clusters
with $\log t_{rh,0}/$yr~$ \gtrsim 9$ and $\Re<0.05$ have
$r_{h,13}/r_{h,0}$ values between 1 and 2.

While stellar evolution and two-body relaxation cause clusters to expand,
galactic tides make clusters shrink in general. A cluster extending farther than
$r_J$ (overfilling; $r_t>r_J$) loses stars outside $r_J$ within a few
dynamical timescales, and this naturally causes the mean size of the cluster
to decrease. Since $r_J \propto R_G (M/M_G)^{1/3}$ where $M_G$ is an
enclosed mass of the Milky Way in a given $R_G$, the size decrease caused by the
galactic tides takes place mostly while the cluster approaches $R_p$.
The cluster re-expands somewhat by two-body relaxation while approaching $R_a$
\citep{bau03}, but its size gradually decreases while repeating orbital
motions. We find that clusters with $0.4 < r_t/r_J < 1$ can also shrink
moderately as a result of the galactic tides even if it underfills, and
clusters initially with $r_t/r_J < 0.4$ (or $\Re < 0.05$; i.e.,
"isolated" GCs) can gradually move into the "tidal" regime as they lose mass
or expand by stellar evolution or two-body relaxation. Figure~\ref{FPresult}
shows that GCs with larger $\Re_0$ are smaller at 13~Gyr for a given
$M_0$ and $R_{G,0}$, as expected.

Among various initial GC parameters, $r_{h,0}$ is the most important parameter
in the size evolution caused by two-body relaxation ($r_{h}/r_{h,0} \propto
M_0^{-1/3}r_{h,0}^{-1}$) and that resulting from galactic tides ($\Re_0
\propto M_0^{-1/3} R_G^{-2/3} r_{h,0}$ for a flat rotation curve). For
this reason, initially small GCs generally expand (by two-body relaxation),
while initially large GCs generally shrink (by the galactic tides) as they
evolve. The size evolution of intermediate GCs is determined by more than one
dynamical effect, and some GCs can even maintain their initial size over their
whole lifetime.

\section{Synthesis of Fokker--Planck Calculations}

As discussed in Section 3, we performed a total of 1152 FP calculations with
different initial cluster conditions in four-dimensional parameter space,
\emph{M}, \emph{$r_{h}$}, \emph{$R_{a}$}, and \emph{e}. The goal of the
present study is to find the initial distribution of these variables that
best describe the observed GCs. For the initial MF model, we adopt a
Schechter function,
\begin{equation}
\setcounter{equation}{1}
{dN(M)\propto{M}^{-\alpha}\exp(-M/M_{s})dM,}
\end{equation}
and for the initial RD model, we use a softened power-law function,
\begin{equation}
\setcounter{equation}{2}
{dN(R_G)\propto{4}{\pi}R_G^{2}dR_G/[1+(R_G/R_{s})^{\beta}].}
\end{equation}
We assume that the initial MF is independent of initial $R_G$. For the sake
of simplicity, we do not parameterize the distribution for $e$, and adopt
the fixed isotropic distributions, i.e., $dN(e)\propto{e}~de$. Unlike $M$,
the $R_G$ of each FP model evolves by oscillating between $R_p$ and $R_a$,
and thus the model RD at 13 Gyr constructed from our population synthesis
may suffer from significant random noise. To reduce this noise, we build a
model RD by summing the probability distributions between $R_p$ and $R_a$ that
are given by the orbital information at 13 Gyr, and we call this a phase-mixed
RD. Hereafter, RDs in this paper refer to the phase-mixed RD.

For initial SDs, we use six distribution models (see Table~\ref{tbl-1}).
Models 1, 2, and 3 represent a Gaussian distribution of $r_h$, $\rho_h$
(mean density within $r_h$), and $\Re$, respectively, implying that the
initial GCs have the preferred initial $r_h$, $\rho_h$, and $\Re$,
with dispersions. The initial $r_h$ of Model 1 does not
correlate with the initial $M$ or $R_G$, while Models 2 and 3 have initial
correlations of $r_{h}\propto{M}^{1/3}{\rho}_{h}^{-1/3}$
and $r_{h}\propto {M}^{1/3}{R_G}^{2/3}{\Re}$. In Models 4, 5, and 6, the
initial $r_h$ is determined by powers of initial $M$ and/or $R_G$. Note that
the power of Model 6 ($r_h \propto M^{0.615}$) corresponds to that of the
mass--size relation derived from the Faber--Jackson relation for early-type
galaxies \citep{fab89,has05,gie10}.

\begin{figure}
\includegraphics[scale=1.0]{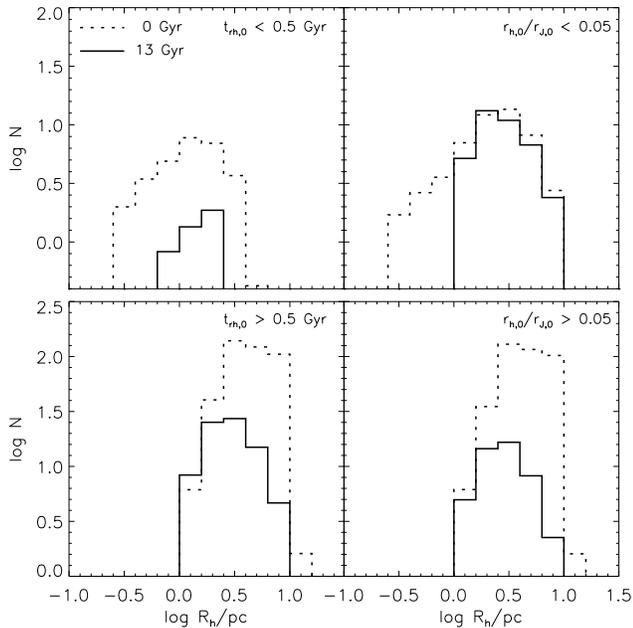}
\caption{Comparison of size distributions at 13~Gyr (solid lines) and at the
beginning (dashed lines) from SD Model 1.
The upper panels are for initially small GCs ($t_{rh,0}<0.5$~Gyr or
$\Re_0<0.05$), and the lower panels are for initially large GCs
($t_{rh,0}>0.5$~Gyr or $\Re_0>0.05$).\label{size}}
\end{figure}

Once the calculations of the 1152 FP models are done, the aforementioned sets of
initial MF, RD, and SD models are used to search for the best-fit parameters
in five to seven dimensional space, depending on the SD models (Models 1--6).
For this, we synthesize our 1152 FP calculations with appropriate weights to
produce a given initial MF, RD, and SD, and find a set of parameters that
best fit the present-day MF, RD, and SD for each of the six SD models. When
finding the best set of parameters for each SD model, we minimize the sum of
${\chi}^{2}$ values from all of the $L$, $R_G$, and $R_h$ histograms, which are
constructed by using eight bins between $10^{4}$ and $10^{5.8}~L_{\odot}$ for
$L$, nine bins between $10^{0}$ and $10^{1.6}$~kpc for $R_G$, and nine bins
between $10^{-0.6}$ and $10^{1.2}$~pc for $R_{h}$, all equally spaced on a
logarithmic scale. Recall that we use dynamical (theoretical) properties
$M$ and $r_h$ for setting the initial distributions, while observable
quantities such as \emph{$L$} and $R_h$ are used for comparing the models and
observations.

\section{Best-fit Initial Distribution of the Galactic Globular Cluster System}

The best-fit parameter sets that minimize the $\chi^2$ values between
observations and our calculations are presented in Table~\ref{tbl-2} for the six
SD models. We examine the characteristics of our best initial MFs, RDs and SDs
in turn.

\subsection{Initial Mass Function}

The best-fit $\alpha$ values for all six SD models are quite low, ranging
between 0.01 and 0.07. The best-fit $\log M_s/$M$_{\odot}$ values for all
six SD models are similar to each other, having values between 5.8 and 5.9.
Note that Schechter functions with such small $\alpha$ values are
similar to log-normal functions, while those of $\alpha \gtrsim2$ are closer to
power-law functions. Thus, our small $\alpha$ values suggest that log-normal
functions better describe the initial MF of the Galactic GC system than
power-law functions (see Figure~\ref{mfrdsd}$a$), and this result is consistent
with the result of Paper I. One way to explain the log-normal-like initial MF is
expulsion of the remnant gas due to star formation in the pre-gas-expulsion
cluster, which can quickly alter a power-law MF into a log-normal-like MF
\citep{par07}.  Another possible mechanism resulting in a rapid change in the
initial MF is the collisions of clusters with dense clouds or other clusters
during the early phase of the galaxy \citep{elm10}.

\subsection{Initial Radial Distribution}

Initial RDs from the best-fit parameter sets for all six SD models have
similar $\beta$ values (4.0--4.5) but a rather wide range of $R_s$ values
(0.3--3.6~kpc), and this is consistent with the result of Paper I ($\beta=4.2$
and $R_s=2.9$~kpc).

Figure~\ref{mfrdsd}$b$ shows that most of the GCs that disrupt before 13~Gyr
are located in the bulge regime ($R_{G,0} < 3$~kpc), and most of the GCs
formed in the bulge do not survive until now.  We find that only 0.1--8.4~\%
of the total GC mass initially inside 3~kpc remains in GCs at 13~Gyr, and
the total stellar mass that escaped from the GCs inside 3~kpc during the
last 13~Gyr amounts to $5 \times 10^7$--$3 \times 10^8$ M$_{\odot}$,
depending on the SD model.

\begin{figure*}
\includegraphics[scale=1.0]{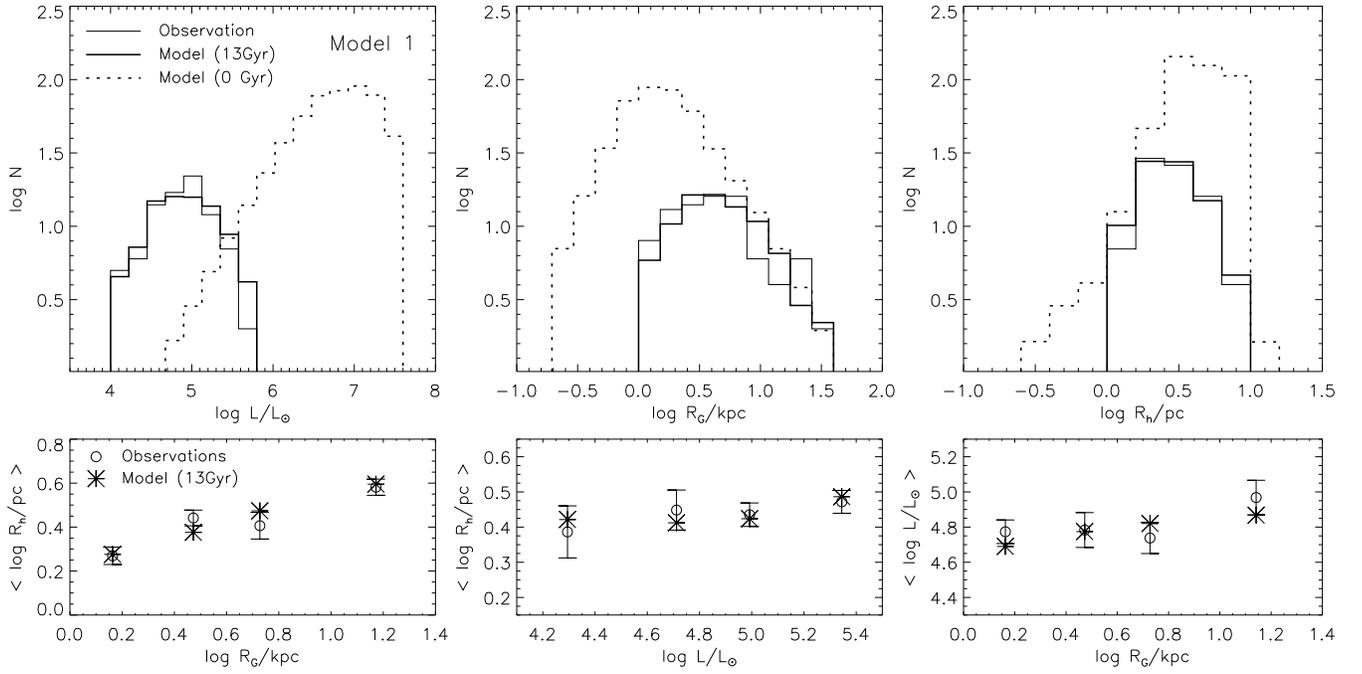} \caption{$L$ (top left),
$R$ (top middle), and $R_h$ (top right) histograms at 13~Gyr (thick solid lines)
for SD Model 1 with the best-fit parameter set.  Also shown together
in the upper panels are the corresponding initial distributions (dashed lines)
and the observed distributions (thin solid lines).
The lower panels show the correlations between $R_h$ and $R_G$ (bottom left),
$R_h$ and $L$ (bottom middle), and $L$ and $R_G$ (bottom right) relationships
for the corresponding best-fit models in the upper panels (asterisks) and
from the observations (open circles).\label{model1}}
\end{figure*}

\begin{figure*}
\includegraphics[scale=1.0]{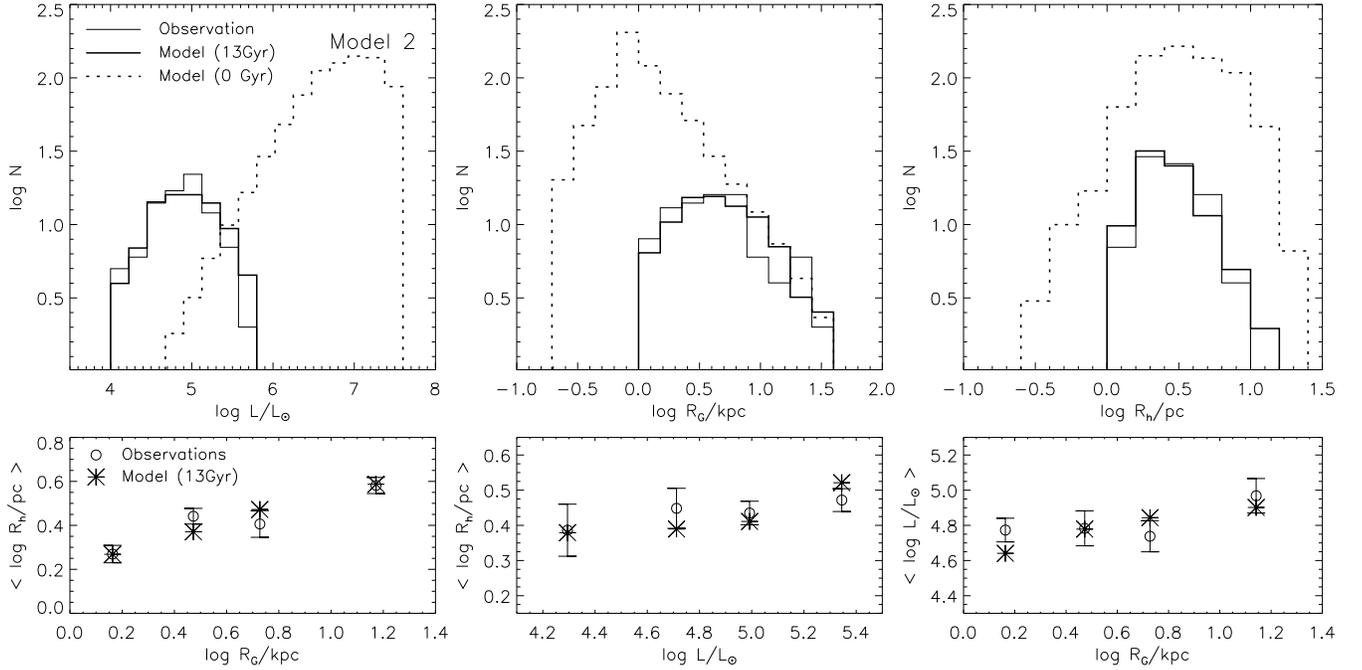} \caption{Same as Figure 5, but
for SD Model 2.\label{model2}}
\end{figure*}

\begin{figure}
\includegraphics[scale=1.0]{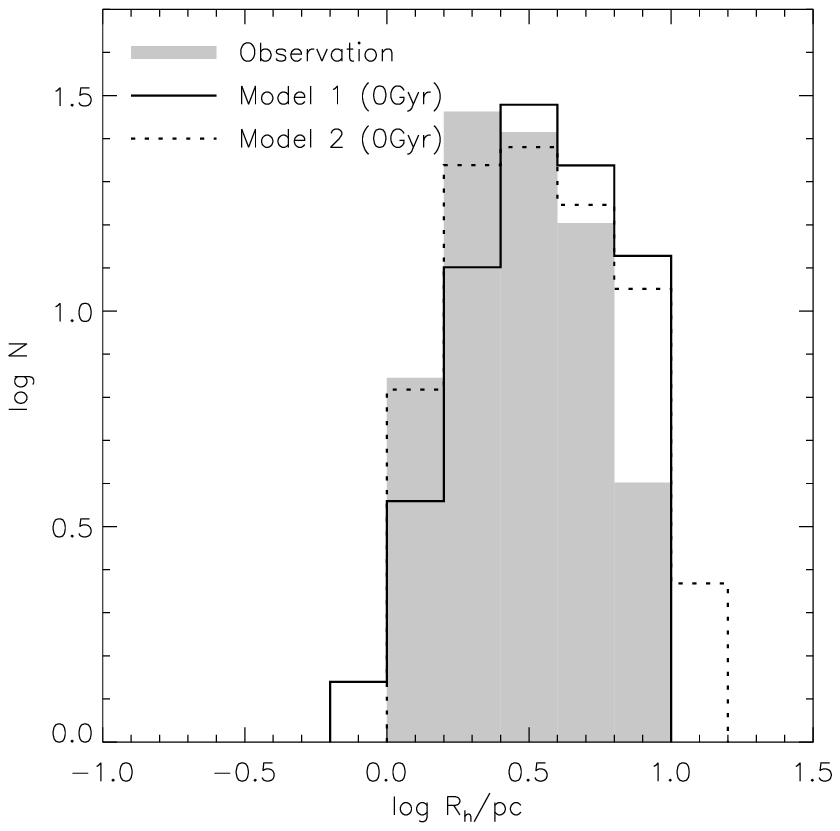}
\caption{Initial SDs of the GCs that survive until 13~Gyr in our best-fit
SD models, Models 1 (thick solid line) and 2 (dashed line).  Also shown
is the currently observed SD (thin solid line with a shaded area). The
overall size of the GCs were larger at birth than now even when only the
surviving GCs are considered.
\label{trace}}
\end{figure}

\subsection{Initial Size Distribution}

The initial SDs from our best-fit parameter sets are of larger dispersion than
the present-day SDs for all six SD models (see Figure~\ref{mfrdsd}$c$). The
initial SDs evolve into the narrower present-day SDs by two main effects:
(1) expansion of GCs with small $r_{h,0}$, which normally have small $t_{rh,0}$
and/or small $\Re_0$, due to two-body relaxation, and (2) shrinkage
(evaporation) of large $r_{h,0}$ GCs, which normally have large $t_{rh,0}$
and/or large $\Re_0$, due to the Galactic tides. Figure~\ref{size} shows
that the SDs of initially small GCs (upper panels) indeed shift to the larger
$r_h$ region and those of initially large GCs (lower panels) shift to the
smaller $r_h$ region after 13~Gyr.

Three $p$-values (significance levels) for $\chi^2$ tests of LFs, RDs, and SDs
are acceptably high, except for Model 3, which has relatively small $\chi^2$
$p$-values for RDs and SDs (see Table~\ref{tbl-2}). However, the high $p$-values
from the $\chi^2$ tests do not necessarily guarantee that the models with the
best-fit parameters restore the observed correlation between $L$, $R_G$,
and $R_h$ as well. Thus, we implement Student's $t$-tests to see if our models
with the best-fit parameters agree with the observed $R_G$ dependence of SDs
(the $R_h$--$R_G$ correlation), the $L$ dependence of SDs (the $R_h$--$L$
correlation), and $R_G$ dependence of LFs (the $L$--$R_G$ correlation).
For the $R_h$--$R_G$ correlation, we calculate $\chi^2$ for the difference of
$\langle\log R_h\rangle$ and $\sigma_{\log R_h}$ between the model and
the observation as follows:
\begin{eqnarray}
\chi^2(\langle\log R_h\rangle) &=& \displaystyle\sum_{j}{\frac{(\langle\log R_{h,o,j}\rangle-\langle\log R_{h,m,j}\rangle)^2}{\sigma^2_{\log R_{h,m,j}}/N_{o,j}}}\nonumber\\
\chi^2(\sigma_{\log{R_h}}) &=& \displaystyle\sum_{j}{\frac{(\sigma_{\log{R_{h,o,j}}}-\sigma_{\log{R_{h,m,j}}})^2}{\sigma^2_{\log R_{h,m,j}}/2N_{o,j}}},\label{eq:xdef}
\end{eqnarray}
where subscripts $o$ and $m$ stand for the observation and the model,
respectively, subscript $j$ represents the equal number $R_G$ bins, and
$\langle\ldots\rangle$ denotes the averaged values. The same calculation is
applied to $R_h$--$L$ and $L$--$R_G$ correlation as well.  We find that
Models 3--6 have $t$-test $p$-values that are too small ($\lesssim1\%$)
for at least one of the $R_h$--$R_G$, $R_h$--$L$, and $L$--$R_G$ correlation.
For this reason, we reject Models 3--6 as being a plausible initial SD
candidate. Hereafter, we call SD models 1 and 2 ``the final best-fit SD
models''.

Figures~\ref{model1} and \ref{model2} show our two remaining best-fit SD
models, a Gaussian distribution of $r_h$ (model 1; $r_{h,c}=6.4$~pc,
$\sigma_{r_h}=2.7$~pc) and a Gaussian distribution of $\rho_h$ (model 2;
$\rho_{h,c}=690~M_{\odot}pc^{-3}$, $\sigma_{\rho_h}=4.6~M_{\odot}pc^{-3}$).
Note that $r_{h,0}$ values are not correlated with the $R_{G,0}$ in either
model. This implies that the $r_{h,0}$ of GCs probably does not depend on the
strength of the galactic tides. Therefore, we interpret the observed,
present-day $R_h$--$R_G$ correlation (see Figure~\ref{obsgcs}$b$) as an
outcome of a preferential disruption of the larger GCs at smaller $R_G$
due to the
Galactic tides.

\section{Discussion}

The typical $R_{h,0}$ value from our final best-fit SD models (Models 1 and 2)
is $\sim4.6$~pc ($r_{h,0}\sim7$~pc), and this is 1.8 times larger than that
of the present-day GCs ($\sim2.5$~pc). This result is rather different
from a recent argument by \citet{bau10} that most GCs were born
compact with $r_{h,0} < 1$~pc.
Our result implies that GCs initially have a rather wide SD, the typical value
of which is similar to that of YMCs in parsec scale, and have evolved to have a
narrower SD with a smaller mean value.

We also find that GCs formation favors a "tidal" environment over an "isolated"
environment. The number of tidal GCs ($\Re_0>0.05$) at 0~Gyr from
our final best-fit SD models is approximately five times larger than that of
isolated GCs ($\Re_0<0.05$). The ratio of tidal to isolated GCs, however,
drastically decreases as GCs evolve because tidal GCs are more easily
disrupted, and this ratio becomes $\sim 0.2$ at 13~Gyr.

Figure~\ref{trace} shows the initial SDs of the GCs that survive until
13~Gyr in Models 1 and 2.  We find that these initial SDs are broader
($\sigma(R_h)=2.1$ and 2.5~pc, respectively) and centered at higher values
($\overline{R_h}$ = 4.1 and 4.0~pc) than the currently observed SD
($\sigma(R_h)=1.2$~pc, $\overline{R_h}=2.5$~pc). Thus, the overall size of
the GCs were larger at birth than now by a factor of $\sim2$ even when only
the surviving GCs are considered.

The initial total masses in GCs ($M_{t,0}$) of the final best-fit SD models are
2.8$\times$${10}^8~M_{\odot}$ (Model 1) and 5.3$\times$${10}^8~M_{\odot}$
(Model 2), and the masses that have left the GCs during the lifetime of the
Galaxy ($\Delta M_t$) are 2.5$\times$${10}^8~M_{\odot}$ (Model 1) and
5.0$\times$${10}^8~M_{\odot}$ (Model 2). These give $\Delta M_t/M_{t,0}$
values of 0.89 and 0.94 for Models 1 and 2, respectively. Our $\Delta M_t$
values are several times larger than previous estimates made by
\citet[4.0--9.5$\times$${10}^7~M_{\odot}$]{bau98},
\citet[5.5$\times$${10}^7~M_{\odot}$]{ves98}, and Paper I
(1.5--1.8$ \times$${10}^8~M_{\odot}$). Our larger $\Delta M_t$ values are
due to the facts that (1) we consider virtually all disruption mechanisms in
the calculations for the dynamical evolution of individual GCs, and (2) we
use more a flexible initial $r_h$ distribution, which can have a relatively
larger fraction of GCs with a large $r_h$ (larger GCs are more vulnerable
to the galactic tide).  Note that $\Delta M_t$ will be larger if one considers
the clusters that have been disrupted in the process of remnant gas expulsion.

We note that contrary to the finding in the present paper, detailed dynamical
modeling of individual clusters shows that at least some of the clusters must
have started with a very small size.  For example, Monte Carlo calculations
by \citet{hg08} and \citet[2011]{gh09} find 0.58 pc,
0.40 pc, and 1.9 pc as best-fit initial $r_h$ values for the observed current
states of M4, NGC 6397, and 47 Tuc, respectively.  These values are several
times smaller than the typical initial $r_h$ found for the Galactic GC system
from our calculations, $\sim 7$~pc.  However, we also note that the Monte
Carlo models used for these three clusters all assume circular cluster orbits
while M4 and NGC 6397 have moderate to high orbit eccentricities (0.82 and
0.34, respectively).  We have performed several FP calculations for these
two clusters and find that consideration of appropriate eccentric orbits
can increase the best-fit initial $r_h$ by a factor of 3--5.

\section{Summary}

We have calculated the dynamical evolution of Galactic GCs using the most
advanced and realistic FP model, and searched a wide parameter space for
the best-fitting initial SD, MF, and RD models that evolve into the present-day
distribution. We found the initial MF of the Galactic GC system is similar to
the log-normal function rather than the power-law function, and the RD of the
GC system undergoes significant evolution inside $R_G = 3$~kpc through
the strong Galactic tides.
We also found that the initial SD of the GC system evolves to narrower
present-day SDs through two effects: shrinkage of large GCs by the galactic
tides and expansion of small GCs by two-body relaxation. The typical initial
projected half-mass radius from the final best-fit model, $\sim 4.6$~pc, is
1.8 times larger than that of the present-day value, $\sim 2.5$~pc. The ratio
of "tidal" GCs to "isolated" GCs is $\sim 5$ at 0 Gyr and decreases down to
$\sim 0.2$ at 13 Gyr.

Since tidal GCs are found to be dominant in the beginning, one might expect
the initial size of the GCs to be correlated with the Jacobi radius, i.e.,
to be a function of the galactocentric radius. However, our final best-fit SD
models (Models 1 and 2) do not seem connected to the galactocentric radius. This
implies that the GC formation process favors a certain size and density,
regardless of the tidal environment. Such a $R_G$-independent initial SD
evolves into a present-day SD, which shows a tight $r_h$--$R_G$ correlation
through evaporation and two-body relaxation.

\acknowledgments
We thank Holger Baumgardt and Mark Gieles for helpful discussion. This work
was supported by Basic Science Research Program (No. 2011-0027247) through
the National Research Foundation (NRF) grant funded by the Ministry of
Education, Science and Technology (MEST) of Korea. This work was partially
supported by WCU program through NRF funded by MEST of Korea
(No. R31-10016). J.S. deeply appreciates Koji Takahashi for the help with
his FP models. S.J.Y. acknowledges support by the NRF of Korea to the Center
for Galaxy Evolution Research and by the Korea Astronomy and Space Science
Institute Research Fund 2011 and 2012. S.J.Y. thanks Daniel
Fabricant, Charles Alcock, Jay Strader, Nelson Caldwell, Dong-Woo Kim, and
Jae-Sub Hong for their hospitality during his stay at Harvard-Smithsonian
Center for Astrophysics as a Visiting Professor in 2011--2012.

\clearpage

\begin{table}
\caption{Initial SD models\label{tbl-1}}
\begin{tabular}{cll}
\tableline\tableline
SD model& Functional form & Parameters \\
\tableline
1&$\mathcal{N}$$(r_{h,c},\sigma^2_{r_h})$\tablenotemark{a}        &$\alpha,M_s,\beta,R_s,r_{h,c},{\sigma}_{r_{h}}$\\
2&$\mathcal{N}$$(\rho_{h,c},\sigma^2_{\rho_h})$\tablenotemark{a}  &$\alpha,M_s,\beta,R_s,\rho_{h,c},{\sigma}_{\rho_h}$\\
3&$\mathcal{N}$$(\Re_c,\sigma^2_{\Re})$\tablenotemark{a}&$\alpha,M_s,\beta,R_s,\Re_c, {\sigma}_{\Re}$\\
4&$r_h=\kappa M^{\lambda}R^{\nu}$ & $\alpha, M_{s},\beta, R_{s}, \kappa, \lambda, \nu$  \\
5&$r_h=\kappa M^{\lambda}$        & $\alpha, M_{s},\beta, R_{s}, \kappa, \lambda$  \\
6&$r_h=\kappa M^{0.615}$         & $\alpha, M_{s},\beta, R_{s}, \kappa$  \\
\tableline
\end{tabular}
\end{table}

\begin{deluxetable}{ccccccclcccccccccc}
\tabletypesize{\footnotesize}
\setlength{\tabcolsep}{0.02in}
\tablecaption{Best-fit parameters for initial GC distributions\label{tbl-2}}
\tablewidth{0pt}
\tablehead{
\colhead{} & \multicolumn{2}{c}{MF} & \colhead{} & \multicolumn{2}{c}{RD} & \colhead{} & \colhead{SD} & \colhead{} & \multicolumn{3}{c}{$p$-values ($\chi^2$ test)} & \colhead{} & \multicolumn{3}{c}{$p$-values ($t$-test)} &\colhead{} &\colhead{}\\
\cline{2-3} \cline{5-6} \cline{8-8} \cline{10-12} \cline{14-16}\\
\colhead{Model}               &
\colhead{$\alpha$}  & \colhead{$\log M_s$} & \colhead{} &
\colhead{$\beta$}   & \colhead{$R_s$}      & \colhead{} &
\colhead{} &
\colhead{} &
\colhead{LF}   & \colhead{RD}  &\colhead{SD} & \colhead{} &
\colhead{$R_h$--$R_G$}   & \colhead{$R_h$--$L$}  &\colhead{$L$--$R_G$}&\colhead{}&\colhead{$M_{t,0}$}\\
\colhead{}               &
\colhead{}  & \colhead{(M$_{\odot}$)} & \colhead{} &
\colhead{}   & \colhead{(kpc)}      & \colhead{} &
\colhead{} &
\colhead{} &
\colhead{($\%$)}   & \colhead{($\%$)}  &\colhead{($\%$)} & \colhead{} &
\colhead{($\%$)}   & \colhead{($\%$)}  &\colhead{($\%$)}&\colhead{}&\colhead{($10^8\,{\rm M_{\odot}}$)}}
\startdata
1 & 0.06 & 5.8 & &4.4 & 1.6 & & $r_{h,c}$=$6.4, \hspace{0.13cm} \sigma_{r_h}$=2.7 & & 72 & 37 & 87 & & 31 & 92 & 48 &&2.8\\
2 & 0.01 & 5.9 & &4.2 & 0.3 & & $\rho_{h,c}$=$690$, $\sigma_{\rho_h}$=4.6         & & 66 & 43 & 63 & & 28 & 42 & 22 &&5.3\\
3 & 0.01 & 5.8 & &4.5 & 3.6 & & $\Re_c$=$0.04, \hspace{0.13cm} \sigma_{\Re}$=0.02 & & 94 & 8  & 11 & & 1  & 79 & 38 &&1.1\\
4 & 0.01 & 5.8 & &4.0 & 1.9 & & $\kappa$=$10^{-3.1}$, $\lambda$=$0.45$, $\nu$=0.3 & & 86 & 27 & 43 & & 3  & 1  & 56 &&1.3\\
5 & 0.07 & 5.9 & &4.0 & 1.7 & & $\kappa$=$10^{-2.0}$, $\lambda$=$0.44$            & & 24 & 28 & 25 & & 1  & 1  & 10 &&1.9\\
6 & 0.05 & 5.9 & &4.4 & 1.9 & & $\kappa$=$10^{-2.9}$                              & & 74 & 21 & 68 & & 15 & 0  & 0  &&5.4
\enddata
\tablecomments{The $p$ value for the $\chi^2$ test ($t$-test) is the
probability of having a $\chi^2$ ($t$) value that is larger than the value
obtained from our $\chi^2$ ($t$) test between the model and the observation,
whose degree of freedom is 8 or 9 (4).  $r_h$ and $r_{h,c}$ are in units of pc,
and $\rho_h$ and $\rho_{h,c}$ are in units of M$_{\odot}/$pc$^3$.}
\end{deluxetable}
\end{document}